
\input phyzzx

\unnumberedchapters
\date={April 1992}    
\Pubnum={\caps UPR--503--T
}

\def\to{\rightarrow}

\titlepage
\title{GRAVITATIONAL
       EFFECTS IN  SUPERSYMMETRIC DOMAIN WALL BACKGROUNDS}

\frontpageskip=0.5\medskipamount plus 0.5 fil
\author{Mirjam Cveti\v c  and  Stephen Griffies}
\address{Department of Physics\break
        University of Pennsylvania\break
        Philadelphia, PA 19104--6396}

\abstract{
A recent study of supersymmetric domain walls in $N=1$
supergravity theories revealed a new class of domain walls
interpolating between
supersymmetric vacua with different  non-positive
cosmological constants.
We classify
three classes of domain wall configurations and study
the geodesic structure of the induced space-time.
Motion of massive test particles in such space-times shows that
these walls are always
repulsive from the anti-deSitter (AdS)
side, while on the Minkowski side test particles
feel no force.
Freely falling
particles far away from a wall in an AdS vacuum experience
a constant proper acceleration, \ie\ they are Rindler particles.
A new coordinate system for discussing AdS space-time
is presented
which eliminates the use of a periodic time-like coordinate.}

\endpage

\REF\VILEN{ A. Vilenkin, Phys. Rep. {\bf 121} (1985) 263.}

\REF\LETT{ M. Cveti\v c, F. Quevedo and S.-J. Rey, Phys. Rev. Lett.
{\bf 63} (1991) 1836. }

\REF\OURS{ M. Cveti\v c, S. Griffies, S.-J. Rey, \sl Static
Domain Walls in $N=1$ Supergravity \rm, UPR-473-T, YCTP-P43-91
(January 1992). }

Domain walls in field theories have been understood for some
time (see \refmark{ \VILEN } for a review).
These objects are inherently relativistic since their
surface tension is precisely equal their surface energy
density.  Such sources of nontrivial tension, or negative pressure,
create a repulsive gravitational field in the sense that
massive test particles are accelerated away from the wall.
In addition,
solving Einstein's equations for an infinitesimally
thin domain wall separating two Minkowski vacua reveals
that the space time metric is time dependent in a choice of
coordinates which reflect the planar symmetry of the wall.

A recent study of domain walls arising in $N=1, D=4$ supergravity
coupled to a chiral superfield discovered a class of vacuum
domain walls with characteristically new features\refmark{\LETT\ ,
\OURS }:
$(1)$ They interpolate between
supersymmetric
vacua with different
non-positive cosmological constants;
 $(2)$ They produce a space-time
metric which is
\sl time-independent, \ie\ the  domain walls are static
 \rm; $(3)$  They extend the notion of vacuum
degeneracy to mean \sl any \rm supersymmetric vacuum in the theory, not
just those with degenerate scalar potential energy.

The purpose of this paper is to classify and
study the structure of the
space-time induced by these domain walls by considering
the motion of massive test particles.  Such motion will fall into
three classes depending on the three possible matter configurations
compatible with the Bogomol'nyi equations.  Comparisons
with non-supersymmetric domain walls will be made as well.
As a byproduct, we introduce a new choice of coordinates
describing an AdS space-time which avoids the use of a periodic
time-like coordinate.

First we review the relevant formulation\refmark{ \OURS }.
By exploiting the
theory's supersymmetry, a Bogomol'nyi bound
for the ADM surface energy density of the planar
(\eg\ in the $(x,y)$ plane)  domain wall
configuration has been derived.
Using the technique of a generalized Nester's form, we
obtain the relation\refmark{\OURS}
$$
\sigma - |C| =
\int_{-\infty}^{\infty}
[-\delta_{\epsilon}\psi^\dagger_i g^{ij} \delta_{\epsilon}\psi_j +
K_{T \bar T}\delta_{\epsilon}\chi^\dagger \delta_{\epsilon}\chi]dz
\geq 0
\eqn\volumeintegral$$
where $\sigma$ is the energy per unit area,
$C$ is the topological charge, $g_{ij}$ is the
metric of the space coordinates\foot{We use the space-time signature
$(+---)$.}
and $K_{T\bar T}>0$
is the second derivative of the K\" ahler potential
for the matter field $T$.
 $\delta_{\epsilon}\psi$ and $\delta_{\epsilon}\chi$
are the supersymmetry variations of the gravitino and
supersymmetric partner of  the matter chiral superfield, respectively.
For supersymmetric bosonic backgrounds,
$\delta_{\epsilon}\psi_{\mu} = \delta_{\epsilon}\chi = 0$.  Thus the
bound on the energy per area $\sigma$ is saturated
by the absolute value of the topological charge $|C|$.
Saturation of this bound yields two coupled
first order differential equations for the matter and
metric (the ``square-root'' of the Euler-Lagrange equations).
We note that in order to ensure a time-like Killing vector,
and thus a well defined notion of energy, we assume from the
start that both the matter and metric are time independent.

The Ansatz for the
space-time metric due to a flat domain wall in the $(x,y)$
plane is
$$ds^{2} = A(z)(dt^{2} - dz^{2})
+B(z)(-dx^{2} - dy^{2}). \eqn\metric
$$
Saturating the Bogomol'nyi bound  \volumeintegral\
yields the following constraints\refmark\OURS
 on the metric $A(z)$, $B(z)$ and
 the scalar component of the scalar superfield $T(z)$:
$$
Im(\partial_{z}T{D_{T}W\over W}) = 0,\eqn\summaryI$$
$$ \partial_z T(z) = -\zeta \sqrt{A}|W|
e^{\kappa K\over 2}K^{T \bar T}
{D_{\overline{T}}\overline{W}\over \overline{W}},\eqn\summaryII
$$
and
$$\partial_{z}A^{-1/2} =
\partial_{z}B^{-1/2} =
 \kappa\zeta|W|e^{\kappa K\over 2}
\eqn\summaryIII$$
where $W(T)$ is the superpotential, $K(T,\bar{T})$ is the K\"ahler
potential, $D_{T}W = W_{T} + \kappa K_{T}W$, $K^{T \bar T} =
(\partial_{T} \partial_{\bar T}K)^{-1}$,
and $\kappa = 8\pi G_{N}$ with $G_N$ being Newton's constant.
$\zeta$ is either
$+1$ or $-1$
and can  change  sign
when and only when $W$ vanishes.

The explicit expression for the ADM mass density (energy per
area or surface tension)
of the supersymmetric domain wall
configuration satisfying  constraints \summaryI\ ,
\summaryII\ and \summaryIII\
is given by
$$
\sigma = |C| \equiv 2 |(\zeta|We^{\kappa K \over 2}|)_{z=+\infty}
-(\zeta|We^{\kappa K \over 2}|)_{z=-\infty}|
\equiv {2 \over \sqrt{3}}\kappa^{-1}|\Delta(\zeta|\Lambda|^{1/2})|
\eqn\admenergy$$
where
 $\Lambda\equiv -3|\kappa W e^{{\kappa K \over 2}}|^{2}$
is the cosmological constant for the supersymmetric
vacuum.\foot{
We define the cosmological constant as follows.
The energy momentum tensor when $T$ is at its vacuum value
($D_{T}W = 0)$ is
$T_{\mu \nu} = -3\kappa|We^{{\kappa K \over 2}}|^{2}g_{\mu \nu}$.
Therefore,
Einstein's equation $R_{\mu \nu} - {1 \over 2}g_{\mu \nu}R
= \kappa T_{\mu \nu}$ can be written
$R_{\mu \nu} - {1 \over 2}g_{\mu \nu}R  = \Lambda g_{\mu \nu}$
with $\Lambda = -3|\kappa W e^{{\kappa K \over 2}}|^{2}$. }
We now comment on eq. \admenergy\ and the eqs. \summaryI\ ,
\summaryII\ , and \summaryIII\ .

It follows from
\admenergy\ that there are
     no static domain walls saturating the Bogomol'nyi bound
that interpolate  between
two supersymmetric vacua with
zero cosmological constant.
 In this case
$W(+\infty)=W(-\infty)=0$
and thus there is no
energy associated
with such a domain wall since $|C|\equiv 0$.
\REF\WALLMETRIC{ A. Vilenkin, Phys Lett. {\bf 133B}(1983) 177;
J. Ipser and P. Sikivie, Phys. Rev. {\bf D30}(1984) 712.}
 This result is in agreement with the results
 of Ref.\refmark{ \WALLMETRIC}, where for infinitesimally
thin domain walls with
asymptotically Minkowski space-times only
time-dependent  metric solutions were obtained.
The result from \admenergy\ implies that static
supersymmetric domain wall  solutions  exist only
if at least one of the vacua is AdS.

Eq. \summaryI\  is a consistency constraint which
specifies the \sl geodesic path \rm
between two supersymmetric vacua
in the
supergravity potential space $e^{{\kappa K \over 2}} W \in \bf C$ when
mapped from the
$z$-axis $(-\infty, + \infty)$.
This geodesic equation
has qualitatively new features
in comparison with
\REF\AT{ E. Abraham and P. Townsend, Nucl. Phys. {\bf 351B}(1991) 313.}
the geodesic equation in the
global supersymmetric case\refmark{ \LETT\ ,\OURS\ ,\AT } .
While in the global case
geodesics
are arbitrary straight lines in the $W-$plane,
the local
geodesic equation in
the limit $\kappa \to 0$ (global
limit of the local supersymmetric theory)
leads to the geodesic equation
$Im({\partial_{z}W\over W})\equiv
\partial_z \vartheta = 0$
where $W$ has been written
as $W(z) =|W|e^{i\vartheta}$.
This in turn implies that
as $\kappa \rightarrow 0$ the local geodesic equation
reduces to the constraint that
  $W$
has to be  a  straight line passing through the
origin;
\ie\ the phase of $W$ has to be
constant mod $\pi$.

This observation in turn implies that
the introduction of gravity
imposes a strong constraint on the type of
domain wall solutions. In particular,
domain wall solutions in the global case interpolating
between vacua
in the $e^{{ \kappa K \over2}}W$ plane
that do not
lie along a straight line passing through the origin
\sl do not \rm have an analogous solution in the local case.
This result is a
manifestation of the singular nature of a perturbation
in Newton's constant; qualitatively different physics
results when $G_{N}\equiv 0$ relative to $G_{N}>0$.

Eq. \summaryII\ for the $T$ field
(the ``square root'' of the equation of motion for $T$)
and eq. \summaryIII\
for the metric (the ``square root'' of the Einstein's equation)
are invariant under $z$ translation as well as
under rescalings of $(A,B)\to \lambda^2 (A,B)$
and $z\to\lambda^{-1}z$.
Additionally,
eq. \summaryII\
implies that $\partial_z T(z)
\rightarrow 0$ as one approaches the supersymmetric minima which are
points where $D_T W=0$, thus indicating
a solution smoothly interpolating between supersymmtric vacua.
In general, the field $T$ reaches the supersymmetric minimum
exponentially fast as a function of $z$.

We now concentrate the equation \summaryIII\
for the metric.
 Our aim is
to classify       all  the
qualitatively different metric
configurations.

First, we wish to emphasize in
\summaryIII\
the singular limit
when gravity is turned off ($\kappa \rightarrow 0$).
As noted earlier, the same  singular limit ($\kappa\to 0$)
is also responsible for the  restrictive
geodesics in the $W$-plane  compared to
 a global theory
which  contains no gravitational information
 ($\kappa =0$).
For $\kappa =0$, the
factors $A$ and  $B$ are constant in the whole space; \ie\ we have
flat space-time everywhere. However, the
moment $\kappa > 0$, $A$ and $B$ vary with $z$.

We can set $A(z)=B(z)$ without loss of generality
which implies that the metric is
conformally flat. Thus, our aim is to study the
nature of the conformal factor $A(z)$.
We classify the  following
three types of static domain wall configurations
which depend on the
nature of the potential of the matter field.

$(I)$ A wall interpolating between
a supersymmetric AdS
vacuum ($|W_{+\infty}|\ne 0$)
and a
Minkowski supersymmetric
vacuum ($|W_{-\infty}|=0$).
{}From \metric\ one sees that  on the Minkowski
side
the conformal factor approaches a
constant which
can be normalized to unity; \ie\
$$A(z)\to 1,\ \   z\to -\infty. \eqn\minm
$$
 On the
AdS side $A(z)$ falls off as
$z^{-2}$  with the
 strength of the fall-off  determined by the strength of
the cosmological constant; \ie\
$$A(z)\to
{ 3\over
{|\Lambda_{+\infty}| z^2}},\ \
z \to +\infty. \eqn\adsm
$$
The surface energy of this configuration as determined from \admenergy\
is:
$$\sigma_{I} = {2 \over \sqrt{3}}\kappa^{-1}|\Lambda_{+\infty}|^{1/2}
\eqn\sigmaI
.$$
Here, the cosmological constant of the supersymmetric
AdS vacuum is $\Lambda_{+\infty} =\break
-3|\kappa
W e^{{\kappa K \over 2}}|^{2}_{+\infty}$.
Note also that
the proper distance on the AdS side,
$d(z) = \int^{z} A^{1\over2}(z')dz'$, grows as ln$z$ since $A \propto
z^{-2}$.
This coordinate system therefore
completely covers the space-time for points on the AdS side of the
wall.

$(II)$ A wall interpolating between
two supersymmetric AdS vacua and where the superpotential
passes through
zero in between.
The
cosmological constant
{\it need not be the same} in both vacua.
The point where
$W=0$
can be chosen at  $z=0$ without loss of generality.
At this  point $\zeta$ changes sign and thus
$\zeta_{+\infty}=-
\zeta_{-\infty}=1$.
 The conformal factor
has the same asympotic behaviour  on both sides of the
domain wall:
$$
A(z) \to
{ 3\over
{|\Lambda_{\pm
\infty}| z^2}},\ \
z \to  \pm \infty\eqn\madsII
$$
while at $z=0$, \ie\ when $W=0$, the conformal factor
levels out, \ie\
$\partial_z A(z)_{z=0}=0$.
In other words   $A(z)$
 has a characteristic  (in general asymmetric)
bell-like
 shape. Again asymptotically it corresponds to the
AdS metric with infinite  proper distance as $z\to\pm\infty$.

The surface energy of this configuration is
$$\sigma_{II} = {2 \over \sqrt{3}}\kappa^{-1}
(|\Lambda|^{1/2}_{-\infty} + |\Lambda|^{1/2}_{+\infty})\eqn\sigmaII.$$

$(III)$ A wall interpolating between
 two AdS vacua, while the superpotential
does \sl not \rm pass through zero. Again, the
cosmological constant
{\it need not be the same} in both vacua.
 In this case, since $|W|$ is never zero,
  $\zeta$ has
the same sign in the whole region, say, +1. Eq.\metric\
 in turn implies that
the conformal factor necessarily
blows up at some coordinate
$z^{*}$.  In general,
the matter field $T$ has long since interpolated
between the two vacua by the time the metric
reaches $z^{*}$.
Thus, the domain wall, defined as the region over which
$T$ moves from one vacuum to another, lies entirely within
the coordinate
region $
z^{*}<z<+\infty$.
The conformal factor has the asymptotic behaviour:
$$\eqalign{
  A(z)&\to
{ 3\over
{|\Lambda_{+\infty}| z^2}},\ \
z \to +\infty \cr
A(z) &\rightarrow {3\over{|\Lambda_{z^*}|(z - z^{*})^2}}
\ \ z\to z^*}
\eqn\singularity$$
The surface energy of this configuration is
$$\sigma_{III} = {2 \over \sqrt{3}}\kappa^{-1}
| |\Lambda|^{1/2}_{z^{*}}
- |\Lambda|^{1/2}_{ + \infty } |.
\eqn\sigmaIII
$$
Note that
the point $z^{*}$ is an infinite proper distance away from any
other point $z > z^{*}$ since $\int dz A^{1/2} \rightarrow
|$ln$(z - z^{*})|$.

In order to understand this singularity as well as the
distinctive $z^{-2}$ behaviour of the conformal factor on the
AdS side of a wall, it is appropriate at this point to
study AdS space-time in a coordinate
system which singles out the $z$ direction.
For this purpose, we consider the metric
$$
ds^{2} = (\alpha z)^{-2}(dt^{2} - dx^{2} - dy^{2} - dz^{2})
\eqn\AdSmetric$$
with $z>0$.  As noted above, this is the form of the metric
on the AdS side of the domain wall when the $T$ field has
reached its supersymmetric vacuum.  In this context, $\alpha$
is related to the cosmological constant by $\Lambda = -3\alpha^{2}$.

Eq. \AdSmetric\ is the form for the metric
describing AdS space-time  where the translational
invariance is broken in the $z$ direction.
The curvature tensor
satisfies the maximally symmetric condition
$R_{\mu\nu\sigma\rho}=\alpha^{2}(g_{\mu\sigma}g_{\nu\rho}
-g_{\mu\rho}g_{\nu\sigma})$, thus ensuring the existence of
10 Killing vectors.  It follows that one
can represent
four dimensional AdS space-time
as the hyperboloid
$\eta_{A B}Y^{A}Y^{B} = \alpha^{-2}$ embedded in the five
dimensional space with flat metric $\eta^{A B}=diag(+---+)$.
We found that the following choice of coordinates
$$
\eqalign{
Y^{0} &= te^{\alpha \tilde{ z}} \cr
Y^{1} &= xe^{\alpha \tilde{z}} \cr
Y^{2} &= ye^{\alpha \tilde{z}} \cr
Y^{3} &= (\alpha)^{-1}sinh(\alpha \tilde{z})
      - {1 \over 2}\alpha e^{\alpha \tilde{z}}(x^{2} + y^{2} -t^{2})  \cr
Y^{4} &= (\alpha)^{-1}cosh(\alpha \tilde{z})
      + {1 \over 2}\alpha e^{\alpha \tilde{z}}(x^{2} + y^{2} -t^{2})
}\eqn\intrinsic$$
yield the metric intrinsic to the surface
$$
ds^{2} = e^{2 \alpha \tilde{z}}(dt^{2} - dx^{2} - dy^{2}) -
d\tilde{z}^{2}.
\eqn\premetric$$
\REF\BRDV{ N. D. Birrell and P. C. W. Davies, Quantum Fields in
Curved Space, 2nd Edition, Cambridge 1989, chapter 5.}
This choice of intrinsic coordinates is motivated from the cosmological
form for
the metric in \sl deSitter \rm space (see, for example
\refmark\BRDV).\foot{We would like to thank F. Wilczek
for pointing  out this analogy  to us.}
By choosing $z = \alpha^{-1}e^{- \alpha \tilde{z}
}$ we recover the
form of the metric in \AdSmetric\ .
\REF\BF{P. Breitenlohner and D. Z. Freedman, Ann. of Phys.
\bf 144 \rm (1982) 249-281.}
The 10 Killing vectors can be written as
$L_{A B} = z_{A}\partial_{B} - z_{B}\partial_{A}$ and satisfy the
$SO(3,2)$ Lie algebra\refmark\BF
$[L_{A B},L_{C D}] = \eta_{BC}L_{AD} - \eta_{AC}L_{BD}
-\eta_{BD}L_{AC} + \eta_{AD}L_{BC}$.

One should note that
this choice of intrinsic coordinates
covers only one-half of the full AdS space-time
since $Y^{3}+Y^{4} > 0$.  By choosing ($Y^{3}, Y^{4}, z) \rightarrow
(-Y^{3}, -Y^{4}, -z)$, we cover the $Y^{3}+Y^{4}<0$ region and have the
metric \AdSmetric\ for $z<0$.
This choice
should be contrasted
with the standard
set of coordinates  respecting spherical symmetry
about an origin which also
completely covers AdS space-time\refmark{\BF}.
In this case the metric has the form
$$
ds^{2} = (\alpha cos\rho)^{-2}(d\eta^{2}
         - d\rho^{2} - sin^{2}\rho (d\theta^{2}
         + sin^{2}\theta d\phi^{2}) )
\eqn\AdSspherical$$
with $0 \le \rho \le \pi/2, 0 \le \theta \le \pi, 0 \le \phi \le 2\pi$.

The time-like coordinate $\eta$ in \AdSspherical\  is periodic.
However, the
coordinates \intrinsic  , in which time
ranges over $-\infty < t < \infty$,  exhibit \sl no \rm
periodic structure.
What we have effectively done in choosing the planar
coordinates \intrinsic\ is to sacrifice a complete covering of
AdS for a  non-periodic time-like
variable.   Observers restriced
to one half of AdS and who use the line element \AdSmetric\
to describe the space-time will see a flat
Minkowski causality structure.  And, as we will see,
the time-like geodesics as seen by these observers will
be hyperbolas in the $t,z$ plane; \ie , freely falling test
particles moving in the $z$ direction are Rindler particles.
\REF\CG{M. Cveti\v c and S. Griffies, work in progress.}
Further study of AdS space-time in this coordinate system is
being pursued\refmark{\CG}.

We return now to the study of the domain wall configurations.
First we note that the
previous discussion of the metric \AdSmetric\
allows for a straightforward interpretation of
singular wall (type $III$) configuration.  What we have is a domain wall
separating two distinct regions of a
generalized AdS space-time possessing a $z$ dependent
cosmological \sl parameter \rm  which \sl never \rm passes
through zero.  The singular point $z^{*}$ corresponds to the
origin $z=0$ in the metric \AdSmetric\ .  On the ``other side''
of $z^{*}$ lives an AdS space-time symmetric to the
$z>z^{*}$ side. Together these two sides completely cover
the whole of the generalzed AdS space-time just as the regions
$z>0$ and $z<0$ in the planar coordinates leading to
\AdSmetric\ cover all of AdS.

The above  discussion of the three types of domain walls
is illustrated by
a simple polynomial form for the superpotential,
a flat K\"ahler manifold: $K = T\bar{T}$,
and a real $T$.
We choose the superpotential
$$
W =
\gamma T[{1\over 5}T^{4} - {1\over 3}T^{2}(a^{2} + b^{2}) + a^{2}b^{2}].
\eqn\superpotential$$
where $\gamma$ is a mass dimension $-2$ parameter which we set to unity
and $a^2$ and $b^2$ are positive  dimension $2$ parameters.
The K\" ahler potential
is chosen to be
$K = T\bar{T}$.
Depending on the value of the parameters $a$ and $b$, the
superpotential \superpotential\
provides us with a set of theories which
accommodate  the above three classes of the domain walls.

 Note that the geodesic  constraint
$Im(\partial_{z}T{D_{T}W\over W}) = 0$ is always satisfied
for $T=\bar{T}$.
The supersymmetric vacuum satisfies
$D_{T}W \equiv W_T+\kappa K_T W =0$, where
$W_{T} = (T^{2} - a^{2})(T^{2} - b^{2})$.
Thus, for $a,b<<1/\sqrt\kappa$,
the supersymmetric vacua
take place
for real values of $T$
near $\pm a, \pm b$.
By continously changing the value of parameters
$a$ and $b$, \superpotential\
provides us with a set of theories which can accommodate
all the three classes of the domain walls discussed  above.
Figures $1,2$ and $3$ display  the conformal factor $A$ for the
these three  classes of the domain walls.
Each example corresponds to a
different choice of the parameters $a$ and $b$, which we
took for
simplicity  to be in the range
$<<1/\sqrt\kappa$.

We now turn to the
study of the
geodesic structure for  the
induced space-time.  To do so, we analyze the motion of
test particles in the background of a supersymmetric domain
wall.
\foot{Note, in general there is a coupling
between test particles and the $T$ field. {\it E. g.},
in string theory where the $T$ field is the modulus field
of the internal compactification, the charged matter field
couples in a very specific way to the $T$ field.
Another interesting example of the coupling is that of the
supersymmetric partner $\chi$ to the $T$ field.  The study of
such test particles is under way\refmark{\CG}. Here we simply
use test particles to map out the geodesics of the space-time. }.
We will find that a test particle
living in an AdS side of a type $I$ or type $II$ wall
is {\it always repelled} from the wall. However, on a Minkowski
side there will be {\it no force}
on the test particle.
We will show
that the
three classes of domain walls discussed above yield three
distinct time-like geodesic motions of test particles.

The motion of massless particles
is trivial since the metric is conformally flat; they
simply define the usual $45^{\circ}$ null rays in a
space-time diagram.  Particles moving in constant $z$
planes will feel no force since the conformal factor is
only a function of the transverse coordinate $z$.  Therefore,
the only interesting geodesics will come from the $1+1$
metric
 $
ds^{2} = A(z)(dt^{2} - dz^{2}).
$
For massive
particles, which live on time-like geodesics, we can
parametrize the motion with the
proper-time element $ds^{2}\equiv d\tau^{2} > 0$.
Rearranging the metric and introducing the conserved energy per
mass $\epsilon\equiv
A  {dt \over d\tau}$
 of the particle yields the equation
$$
({dz \over dt})^{2} + {A \over \epsilon^{2}} = 1.
\eqn\conservation$$
On a time-like geodesic,
$0 \le ({dz / dt})^{2} < 1$, and so
the turning point, \ie\
 $v \equiv dz/dt = 0$, of the motion
is where
${A / \epsilon^{2}} = 1$.

A convenient way to understand massive particle motion is to
consider a particle with a given initial coordinate velocity
$v_{o}$ at some coordinate $z_{o}$;
from \conservation\
$\epsilon$ for such a particle is
$\epsilon^{2} = A(z_{o})(1-v^{2}_{o})^{-1}$.  Equation \conservation\
can be thought of as the conservation of energy with an
effective potential $V(z)\equiv (1-v^2)
 = {A(z) \over A(z_{o})}(1-v_{o}^{2})$.
Again, points where $V(z) = 1$ are turning
points.

The three classes of domain walls discussed above yield three
time-like geodesics.

$(I)$
Consider a particle
with initial coordinate velocity $-v_{o} \le 0$
moving towards a
Minkowski-AdS wall (class $I$; see Figure 1)
centered at the origin and let the
approach be from the AdS side ($z>0$).
There will be a turning point if the initial velocity of the
test particle satisfies
$v_{o}^{2} \le 1 - A(z_{o})\equiv v_{c}$.
Particles with $v^{2}_{o}$ above the critical
value $1-A(z_{o})$ will penetrate the wall after
a period of slowing down and move at constant
velocity in the Minkowski side.
Figure 4  shows the phase plane (${v={dz \over dt}}$ vs. $z$) for
particles incident on a Minkowski-AdS wall from the $z>0$ AdS side.
Note the particles asymptote to the null ray $v={dz \over dt} = 1$
as $z \rightarrow \infty$  since $A(z)\to 0$ as $z\to +\infty$.

$(II)$
A wall separating two AdS space-times with $W$ passing through
zero (class $II
$; see  Figure 2)
will have turning points on both sides if an initial
velocity of a test particle incident on the wall
is below the critical velocity.
A particle with enough energy can pass
from one side to the other after having slowed down in
the transition region.

$(III)$
Particles incident
on the singular wall configuration (class $III
$; see Figure 3)
from the $z > z^{*}$ side will
always reach a turning point
no matter how much initial energy it has.
One
should not think of this behaviour as being due to an
infinitely repulsive domain wall.  Indeed, as discussed above,
the wall region (around $z=0$)
is well away from $z^{*}$ which
is actually
 an infinite proper distance away.
This behaviour is
a reflection of the repulsive nature of AdS space-time.
A similar behaviour is seen
in
test particles moving radially outwards
in pure
 AdS  space whose metric is written in the spherically symmetric form
\AdSspherical\  as well as particles directed toward $z = 0$ in the
planar form \AdSmetric\ .
Both of these points correspond to proper distance infinity.
In this
sense AdS space
can be thought of as a contracting
space-time since
particles can never reach proper radial $\infty$ ($\rho=\pi/2$)
in the spherical case or proper $-\infty$ ($z=0$) in the planar case.
We note that a particle in the planar coordinates \AdSmetric\
can, however,
reach proper distance
$+\infty$.

Another way to understand the repulsive nature of these space-times
to calculate the force on a test particle which
has a fixed position $z$ (also known as a fiducial observer).
This force can be obtained through the
geodesic equation
$ p^{\alpha}p^{\beta}_{; \alpha} = mf^{\beta} $
with $p^{\alpha} = m{dx^{\alpha} \over d\tau}$.\foot{
Our convention for the gravitational covariant derivative is
$p^{\alpha}_{; \mu} = {d p^{\alpha} \over d x^{\mu}}
+ \Gamma^{\alpha}_{\mu \nu}p^{\nu}$ with
$\Gamma^{\alpha}_{\mu \nu} = {1 \over 2} g^{\alpha \beta}
(g_{\beta \mu , \nu} + g_{\beta \nu , \mu} - g_{\mu \nu , \beta})$. }
The gravitational force acting on the fiducial observer is
$$ f^{\beta} = ( 0, 0, 0,  -{m \over 2} A^{-2} \partial_{z}A).
\eqn\force$$
For a metric which falls off as on the AdS side of a wall, this
force is  directed
\sl towards \rm the AdS vacuum (e.g. $z=+\infty$ in the type $I$
wall depicted in figure $1$).
The magnitude of the
acceleration
is given by
$$ |a|^{2} \equiv |f_{\alpha}f^{\alpha}|/m^{2}
= ({1 \over 2}{ \partial_{z}lnA \over A^{1/2} })^{2} =
  (\kappa |W|e^{{\kappa K\over 2}})^{2}
\eqn\acceleration$$
where eq. \summaryIII\ was used in the last equality.
For fiducial observers in the region where $T$ is
essentially at its vacuum value; \ie\ far away from the wall,
     the proper acceleration has the
constant magnitude
$|a|^2  = |{\Lambda_{\pm\infty}/3}| $.
Additionally, integration of \conservation\ yields the
hyperbolic world line for freely falling test particles
$z^{2}-t^{2}=|{\Lambda_{\pm \infty}/
 3}|^{-1}\epsilon^{-2}$.  Therefore, a
fiducial observer situated far away from a type $I$ or $II$ wall
in a
$\Lambda_{\pm\infty} \ne 0$ region will feel a constant acceleration
$|{\Lambda_{\pm\infty}/
 3}|^{1/2}$ directed away from the wall as well as
see test particles of energy per mass $\epsilon$ moving
away from
the wall with acceleration $|{\Lambda_{\pm\infty}/ 3}|^{1/2}\epsilon$.
\REF\RINDLER{ W. Rindler, \sl Essential Relativity \rm,
Springer-Verlag 1979. }
Thus, asymptotically in the
AdS space-times test particles moving in the $z$ direction  are Rindler
particles\refmark{\RINDLER}
whose constant proper acceleration is proportional to the
square root of  the cosmological constant,
while on the Minkowski side they experience no gravitational force.

This result should be contrasted with the
observation in
Ref.\refmark\WALLMETRIC, where  infinitesimally
thin reflection symmetric
domain walls with
asymptotically Minkowski space-times
always repell
the fiducial observer
with a constant acceleration $\kappa\sigma/4$.
Here, $\sigma$ is the energy per unit area of the
domain wall.
\foot{
Domain walls which separate two Minkowski vacua yet satisfy the
nonstandard relation  $\sigma = 2\tau$, where $\tau$ is the
surface tension of the wall, produce \sl no \rm gravitational force on
test particles.
\REF\IPSER{ J. Ipser, Phys Rev. \bf D30 \rm (1984) 2452.}
Walls of isotropically and uniformly distributed
cosmic strings produce such an equation of state\refmark\IPSER .}
Note that these domain walls always  produce
a time dependent metric.
In our case everything is static.
In particular, for the type $(I)$ domain walls
interpolating between AdS and Minkowski  space-times,
the asymptotic acceleration
on the AdS side can be written as
$a=\kappa\sigma_I/2$, where  $\sigma_I$ is the energy per
unit area of the domain wall $(I)$ defined in \sigmaI\ .
However, on the Minkowski side $a\to 0$.
For the type $II$ domain wall
when the potential has $Z_2$ symmetry (see Fig. 2:
both  AdS  vacua are degenerate),
the energy per unit area
\sigmaII\ is $\sigma_{II}= 4\kappa^{-1}
|{\Lambda_{\pm\infty}/
 3}|^{1/2}$
and the fiducial observer is repelled on both sides of the
domain wall with the same acceleration  $a_{\pm \infty}\to
\kappa \sigma_{II}/4$ which resembles
remarkably the form for the acceleration for the
 domain walls discussed in Ref.\refmark\WALLMETRIC\ .
In our case the domain wall also respects the
$Z_2$ symmetry, however, it is completely static
and its repulsive nature is due to
the AdS nature of the asymptotic space-time.

We would like to thank
S.-J. Rey
for many enlightening discussions and enjoyable collaboration.
We also benefitted from discussions
with  D. Muraki, E. Weinberg, and
F. Wilczek.
The research is  supported in part by the
U.S. DOE Grant DE5-22418-281, by
the NATO Research Grant
No. 900-700 (M.C.),
and by the junior faculty
SSC fellowship (M.C.).

\endpage

\refout

\endpage

{\bf Figure Captions}

\noindent{Figure $1$:
Type $(I)$
conformal factor $A(z)$ for a space-time with $\Lambda_{-\infty}
= 0$
(Minkowski: $z<0$) separated by a domain wall from a space-time
with $\Lambda_{+\infty}
 < 0$ (AdS: $z>0$).  The wall, \ie\
 the
region over which the matter field $T$ changes is centered
at $z=0$  and has thickness
$\approx 200$ in $ \sqrt\kappa$ units.
The superpotential \superpotential\ has parameters $a^{2}=0,
b^{2}=0.1$ and $T$
interpolates between  $T_{-\infty}= 0 = a$
and $T_{\infty}= .318 \approx b$
. }

\
\

\noindent{
Figure $2$:  Type $(II)$
Conformal factor $A(z)$ for a space-time with negative
cosmological constant separated by a domain wall from its mirror image
(\ie\ a $Z_{2}$ configuration).  The wall is centered at $z=0$ and has
thickness $\approx 200$ in $\sqrt\kappa$ units.
The superpotential
\superpotential\  has parameters $a^{2}=.025, b^{2}=0.1$.
and $T$
interpolates between  $T_{\mp\infty}= \pm .1598 \approx \pm a$
. }

\
\

\noindent{
Figure $3$: Type $(III)$
conformal factor $A(z)$ for a space with negative
cosmological constant
separated by a domain wall from a space with a different
negative cosmological constant.  The superpotential $W$ never passes
through a zero as $T$ interpolates from one vacuum to another.
The domain wall is centered at $z=0$ and
has thickness $\approx 200$
where   $z$q is measured in $\sqrt\kappa$ units.
The singularity is at $z^{*}
\approx -5600$.
The superpotential
\superpotential\ has parameters $a^{2}=.025, b^{2}=0.1$
and $T$
interpolates between  $T_{-\infty}= .315 \approx b $
and $T_{\infty}= .160 \approx a$
.}

\
\

\noindent{
Figure $4$: Phase plane ($v\equiv
{dz \over dt}$ vs. $z$) for massive test
particles incident on a Minkowski-AdS (class $(I)$
) wall from
the AdS side.  Particles which start at $z_{o}$ with an initial
velocity $v_{o}$ will reach a turning point if $v_{o}^{2} \le
v_{c}^{2} \equiv 1-A(z_{o})$.
Particles with a turning point will return to $z_{o}$ with
the same velocity $v_{o}$ now directed away from the wall and
will continue to accelerate toward the speed of light.  Particles
with $v_{o}^{2} > v_{c}$ will penetrate the wall region after
slowing down and continue on in the Minkowski side with a
constant velocity.}
\
\

\end